\documentclass[a4paper]{jpconf}
\usepackage{amsmath}
\usepackage{graphicx}
\usepackage{latexsym}  
\usepackage{amsfonts}  
\usepackage{amssymb} 
\usepackage{color,mathtools}
\usepackage{subfig,cancel}
\usepackage{dsfont,xcolor}
\usepackage{url}
\usepackage[square,sort&compress,numbers]{natbib}
\usepackage{verbatim}
\usepackage[T1]{fontenc}
\usepackage[utf8x]{inputenc}

\newcommand{\be}{\begin{equation}}  
\newcommand{\ee}{\end{equation}}
\newcommand{\beq}{\begin{eqnarray}}  
\newcommand{\eeq}{\end{eqnarray}}  

\begin{document}

\def\hc{\text{h.c.}}
\def\ud{{\mathrm{d}}}
\def\udl{\stackrel{\leftarrow}{\ud}}
\def\w{{\omega}}
\def\im{{\mathrm{i}}}
\def\ex{{\mathrm{e}}}
\newcommand{\ket}[1]{| #1 \rangle}
\newcommand{\bra}[1]{\langle #1 |}
\newcommand{\ev}[2]{\langle #1 | #2 \rangle}
\newcommand{\mel}[3]{\langle #1 | #2 | #3 \rangle}
\def\intw{{\int\hspace{-2pt}\frac{\ud \omega}{2\pi}}}
\def\intwp{{\int\hspace{-2pt}\frac{\ud \omega'}{2\pi}}}
\def\intwf{{\int\hspace{-2pt}\frac{\ud \omega}{2\pi}}f(\omega - \mu)}
\renewcommand\Im{\operatorname{Im}}
\renewcommand\Re{\operatorname{Re}}

\def\tb{\bar{t}}  
\def\zb{\bar{z}}  
\def\tgb{\bar{\tau}}  
\def\bC{\mbox{\boldmath $C$}}  
\def\bG{\mbox{\boldmath $G$}}  
\def\bH{\mbox{\boldmath $h$}}  
\def\bK{\mbox{\boldmath $K$}}  
\def\bM{\mbox{\boldmath $M$}}  
\def\bN{\mbox{\boldmath $N$}}  
\def\bO{\mbox{\boldmath $O$}}  
\def\bQ{\mbox{\boldmath $Q$}}  
\def\bR{\mbox{\boldmath $R$}}  
\def\bS{\mbox{\boldmath $S$}}  
\def\bT{\mbox{\boldmath $T$}}  
\def\bU{\mbox{\boldmath $U$}}  
\def\bV{\mbox{\boldmath $V$}}  
\def\bZ{\mbox{\boldmath $Z$}}  
\def\unit{\mbox{\boldmath $1$}}
\def\bcalS{\mbox{\boldmath $\mathcal{S}$}}  
\def\bcalG{\mbox{\boldmath $\mathcal{G}$}}  
\def\bcalE{\mbox{\boldmath $\mathcal{E}$}}  
\def\bgG{\mbox{\boldmath $\Gamma$}}  
\def\bgL{\mbox{\boldmath $\Lambda$}}  
\def\bgS{\mbox{\boldmath $\mathit{\Sigma}$}}  
\def\a{\alpha}  
\def\b{\beta}  
\def\g{\gamma}  
\def\G{\Gamma}  
\def\d{\delta}  
\def\D{\Delta}
\def\eps{\epsilon}
\def\e{\mathbf{e}}
\def\z{\zeta}  
\def\h{\eta}  
\def\th{\theta}  
\def\l{\lambda}  
\def\L{\Lambda}  
\def\t{\tau}  
\def\f{\phi}  
\def\vf{\varphi}  
\def\F{\Phi}  
\def\c{\chi}  
\def\Q{\Psi}  
\def\q{\psi}  
\def\heff{h_{\text{eff}}}
\def\heffd{h_{\text{eff}}^{\dagger}}
\def\hypf{{}_2{F}_1}

\title{Time-dependent Landauer--B{\"u}ttiker formalism for superconducting junctions at arbitrary temperatures}

\author{Riku Tuovinen$^1$, Robert van Leeuwen$^{1,4}$, Enrico 
Perfetto$^{2}$ and Gianluca Stefanucci$^{2,3,4}$}

\address{$^1$ Department of Physics, Nanoscience Center, FIN 40014,
University of Jyv\"askyl\"a, Jyv\"askyl\"a, Finland}
\address{$^2$ Dipartimento di Fisica, Universit\`a di Roma Tor
Vergata, Via della Ricerca Scientifica 1, I-00133 Rome, Italy}
\address{$^3$ Laboratori Nazionali di Frascati, Istituto Nazionale di
Fisica Nucleare, Via E. Fermi 40, 00044 Frascati, Italy}
\address{$^4$ European Theoretical Spectroscopy Facility (ETSF)}

\ead{riku.m.tuovinen@jyu.fi}

\begin{abstract}
We discuss an extension of our earlier work on the time-dependent Landauer--Büttiker formalism for noninteracting electronic transport. The formalism can without complication be extended to superconducting central regions since the Green's functions in the Nambu representation satisfy the same equations of motion which, in turn, leads to the same closed expression for the equal-time lesser Green's function, i.e., for the time-dependent reduced one-particle density matrix. We further write the finite-temperature frequency integrals in terms of known special functions thereby considerably speeding up the computation. Simulations in simple normal metal -- superconductor -- normal metal junctions are also presented.
\end{abstract}

\section{Introduction}\label{sec:intro}
The process of Andreev reflection\cite{Andreev1964} (AR) occurring at the interface between a normal metal (N) and a superconductor (S) is of great importance with applications in spintronics and quantum computing. An incoming electron from N to S produces a Cooper pair in S and a reflected hole in N\cite{Cooper1956,BCS1,BCS2}. In an NSN junction normal metal electrodes are spatially separated by a superconducting central region and an entangled electron--hole pair can be transported. This can be seen when the junction separation is of the order of the superconducting coherence length for the studied material, and when the incident electron energies are less than the superconducting gap for the AR process to occur\cite{PhysRevLett.70.1862}.

The quantum transport problems are typically time dependent; there is no guarantee that the system would in an instant relax to a steady-state configuration once the junction is ``switched on'' (as in connecting different devices or driving them out of equilibrium by an external perturbation). In contrast, there are transient effects depending on, e.g., the system's geometry\cite{Khosravi2009,Perfetto2010,Vieira2009,Wang2015}, its predisposition to external perturbations\cite{Kurth2010,Arrachea2010,Ness2011,Arrachea2012}, the physical properties of the transported quanta and their mutual interactions\cite{Wijewardane2005,Verdozzi2006,Myohanen2008,Uimonen2011,Myohanen2012ic,Latini2014}. Even if the transport mechanisms were discussed in an idealized noninteracting setting, it is therefore important to consider a fully time-dependent description of the studied processes.

The Landauer--Büttiker formalism is simple to understand as it relates to an intuitive physical picture of charge transport in a multiterminal junction\cite{landauer,buttiker}. Including the transient description to the formalism by studying the nonequilibrium Green's function approach does not complicate the final result\cite{caroli1,caroli2,cini,meir-wingreen,jauho,perfetto}; the physical picture is still clear and intuitive as different features of the transport setup can be directly linked to the time-dependence\cite{svlbook,Tuovinen2013,Tuovinen2014}. In this paper, we present an extension to earlier results for both superconducting junctions and arbitrary temperatures (Sec.~\ref{sec:background}). We present a formula for the time-dependent one-particle reduced density matrix (TD1RDM) as such since it is a closed, analytic expression which can readily be implemented for numerical model systems. Further details of the derivation are to be found in another work\cite{Tuovinen2015}. In Sec.~\ref{sec:results} we illustrate the features of the formula by studying transients in simple NSN junctions.

\section{Background and Nambu representation}\label{sec:background}
We consider a quantum transport setup similar to one studied in the previous volume of this conference series\cite{Tuovinen2013}. In this setup, a noninteracting central region is connected between metallic leads, and the Hamiltonian takes the form
\beq\label{eq:hamiltonian}
\hat{H} & = & \hat{H}_{\text{leads}} + \hat{H}_{\text{central}} + \hat{H}_{\text{coupling}}\nonumber \\
& = & \sum_{k\alpha\sigma}\left[\epsilon_{k\alpha}+\theta(t)V_\alpha\right]\hat{c}_{k\alpha\sigma}^\dagger \hat{c}_{k\alpha\sigma} + \sum_{mn\sigma} T_{mn}\hat{c}_{m\sigma}^\dagger \hat{c}_{n\sigma} + \sum_{mk\alpha\sigma}\left(T_{mk\alpha}\hat{c}_{m\sigma}^\dagger \hat{c}_{k\alpha\sigma} + \hc\right) . 
\eeq
The operators $\hat{c}^{(\dagger)}$ annihilate (create) electrons from (to) a region specified by the subscript indices: $k\alpha$ is the $k$-th basis element of the $\alpha$-th lead, $m,n$ label the basis elements of the central region, and $\sigma\in\{\uparrow,\downarrow\}$ is a spin-$\frac{1}{2}$ index. These operators moreover obey the fermionic anticommutation relations $\{\hat{c}_{x\sigma},\hat{c}_{y\sigma'}^\dagger\}=\delta_{xy}\delta_{\sigma\sigma'}$. The Hamiltonian structure is determined by the single-particle levels in the leads $\epsilon_{k\alpha}$ and the tunneling matrices $T$ between the states of the central region ($T_{mn}$) and between the states of the central region and the leads ($T_{mk\alpha}$). The system is driven out of equilibrium for times $t>0$ by a sudden shift of the lead energy levels by $V_\alpha$. In addition to a sudden bias, it is also possible to include time-dependent bias profiles without complicating the following derivations\cite{Ridley2015}. In order to describe a superconducting island, we will now add a pairing field operator $\hat{\varDelta}$ to the Hamiltonian of the central region by\cite{Jauho2000,Stefanucci2010}
\be
\hat{H}_{\text{central}} \to \sum_{mn\sigma} T_{mn}\hat{c}_{m\sigma}^\dagger \hat{c}_{n\sigma} + \sum_m \varDelta_m \hat{c}_{m\uparrow}^\dagger \hat{c}_{m\downarrow}^\dagger + \sum_m \varDelta_m^* \hat{c}_{m\downarrow} \hat{c}_{m\uparrow} .
\ee
The one-electron Green's function in the above setup can be defined via the \emph{Nambu spinor} $\underline{\hat{\varPhi}}_m = (\hat{\varPhi}_{m}^1 , \hat{\varPhi}_{m}^2)^{T} = (\hat{c}_{m\uparrow} , \hat{c}_{m\downarrow}^\dagger)^{T}$, which obeys the anticommutation relation in a tensor product sense $\{\hat{\varPhi}_m^\mu,\hat{\varPhi}_n^\nu\} = \delta_{mn}\delta^{\mu\nu}$, as a contour-ordered product\cite{Zeng2003,Stefanucci2010}
\be\label{eq:green}
\underline{G}_{rs}(z,z') = -\im\langle \mathcal{T}_\gamma[\underline{\hat{\varPhi}}_r(z) \otimes \underline{\hat{\varPhi}}_s^\dagger(z')] \rangle
\ee
where the contour-ordering operator $\mathcal{T}_\gamma$ is taken for the variables $z,z'$ on the Keldysh contour $\gamma$\cite{svlbook}. When the product in Eq.~\eqref{eq:green} is expanded, the elements in the resulting $2\times 2$ \emph{Nambu matrix} are the normal and the anomalous components of the Green's function\cite{Nambu1960,Stefanucci2010PNGF}. As already denoted above, we put an underline for quantities in the Nambu space. The matrix elements in the Green's function label the transport setup in the following block form [notice that the indices $r,s$ in Eq.~\eqref{eq:green} may belong to any block and that the number of leads is arbitrary]
\be\label{eq:hamiltonian-block}
\underline{\bH} = \begin{pmatrix}\underline{h}_{11} & 0 & \cdots & \underline{h}_{1 C} \\ 0 & \underline{h}_{22} & \cdots & \underline{h}_{2C} \\ \vdots & \vdots & \ddots & \vdots \\ 
\underline{h}_{C1} & \underline{h}_{C2} & \cdots & \underline{h}_{CC} \end{pmatrix} \ ; \qquad \underline{\bG} = \begin{pmatrix}\underline{G}_{11} & \underline{G}_{12} & \cdots & \underline{G}_{1 C} \\ \underline{G}_{21} & \underline{G}_{22} & \cdots & \underline{G}_{2C} \\ \vdots & \vdots & \ddots & \vdots \\ 
\underline{G}_{C1} & \underline{G}_{C2} & \cdots & \underline{G}_{CC} \end{pmatrix}
\ee
with 
\beq
(\underline{h}_{\alpha \alpha'})_{kk'}(t) & = & \begin{pmatrix}[\epsilon_{k\a}+\theta(t)V_\a]\delta_{\alpha\alpha'}\delta_{kk'} & 0 \\ 0 & -[\epsilon_{k\a}+\theta(t)V_\a]\delta_{\alpha\alpha'}\delta_{kk'} \end{pmatrix}, \label{eq:halpha}\\
(\underline{h}_{CC})_{mn} & = & \begin{pmatrix}T_{mn} & \varDelta_{m}\delta_{mn} \\ \varDelta_{m}^*\delta_{mn} & -T_{mn}\end{pmatrix},\label{eq:hc} \\
(\underline{h}_{C\alpha})_{mk\alpha} & = & \begin{pmatrix}T_{mk\a} & 0 \\ 0 & -T_{mk\a}\end{pmatrix} , \quad (\underline{h}_{\alpha C})_{k\alpha m} = \begin{pmatrix}T_{k\a m} & 0 \\ 0 & -T_{k\a m}\end{pmatrix}
\eeq
for the leads, central region and couplings, respectively. It is important to notice that even though only the central region is superconducting, all the blocks in the Hamiltonian are written in the form of Bogoliubov--de Gennes\cite{Bogoliubov1958,deGennes1966}. Including the pairing field in the Hamiltonian of the central region adds no extra complication to the evolution of the Green's function\cite{Tuovinen2015}; the only difference, compared to the earlier work in Refs.~\cite{svlbook,Tuovinen2013,Tuovinen2014}, is in the interpretation of the matrices in Nambu space. It is also possible to include non-local pairing field $\varDelta_{mn}$ with arbitrary spin-coupling leading to $4$-component Nambu spinors which includes the possibility to study also Majorana fermions\cite{Flensberg2012,Tuovinen2015}. The Hamiltonian and the Green's function are connected via the equation of motion (with the boundary condition that the Green's function is antiperiodic along the contour, i.e., the Kubo--Martin--Schwinger boundary conditions\cite{Kubo1957,Martin1957,svlbook})
\be\label{eq:eom}
\left[\im\frac{\ud}{\ud z}\underline{\unit} - \underline{\bH}(z)\right]\underline{\bG}(z,z') = \delta(z,z')\underline{\unit}
\ee
and the corresponding adjoint one. We describe the leads within the wide-band approximation (WBA), where the electronic levels of the central region are in a narrow range compared to the lead bandwidth which gives for the retarded embedding self-energy
\be\label{eq:sigmar}
\underline{\varSigma}_{\a,mn}^\text{R}(\w) = \sum_k (\underline{h}_{C\a})_{mk\a}\frac{1}{\w-\underline{\eps}_{k\a}-\underline{V}_\a+\im\eta} (\underline{h}_{\a C})_{k\a n} \approx -\im\underline{\varGamma}_{\a,mn}/2 ,
\ee
where the bandwidth matrices satisfy $\underline{\varGamma} = \sum_\a \underline{\varGamma}_\a$. For the lead Green's function between the coupling Hamiltonians in Eq.~\eqref{eq:sigmar} the structure is similar to that of Eq.~\eqref{eq:halpha}. Approximating the embedding self-energy this way, as a purely imaginary constant, closes the equation of motion~\eqref{eq:eom}, and it can then be solved analytically\cite{svlbook,Tuovinen2013,Tuovinen2014}. As the solution we get the lesser Green's function (in region $CC$) in the equal-time limit and the TD1RDM by
\beq\label{eq:green2}
(\underline{\rho}_{CC})_{mn}(t) = -\im (\underline{G}_{CC}^<)_{mn}(t,t) & = & -\im\begin{pmatrix}(G^<_{CC,\uparrow})_{mn}(t,t) & (-F^>_{CC})_{nm}(t,t) \\ (\bar{F}^<_{CC})_{mn}(t,t) & (-G^>_{CC,\downarrow})_{nm}(t,t)\end{pmatrix} \nonumber \\
& = & \begin{pmatrix}\langle\hat{c}_{n\uparrow}^\dagger(t)\hat{c}_{m\uparrow}(t)\rangle & \langle\hat{c}_{n\downarrow}(t)\hat{c}_{m\uparrow}(t)\rangle \\ \langle\hat{c}_{n\uparrow}^\dagger(t)\hat{c}_{m\downarrow}^\dagger(t)\rangle & \langle\hat{c}_{n\downarrow}(t)\hat{c}_{m\downarrow}^\dagger(t)\rangle\end{pmatrix}
\eeq
where the normal ($G^{\lessgtr}_\sigma$) and anomalous ($F^{\lessgtr}$) components\cite{Stefanucci2010} follow by expanding the product in Eq.~\eqref{eq:green}. It is also possible to solve the equations of motion analytically for the two-time Keldysh components of the Green's function; this offers the possibility to extract not only densities and currents but other physical quantities such as noise from the solution\cite{Ridley2015}.

Expressed in the left eigenbasis, $\bra{{\varPsi}^{\text{L}}}\underline{h}_{\text{eff}} = {\eps}\bra{{\varPsi}^{\text{L}}}$ (the eigenvalues $\eps$ are in general complex), of the nonhermitian effective Hamiltonian $\underline{h}_{\text{eff}} = \underline{h}_{CC} - \im\underline{\varGamma}/2$ the matrix elements of the TD1RDM take the explicit form\cite{Tuovinen2014,Tuovinen2015}
\beq
&  & \langle {\varPsi}_j^{\text{L}} | \underline{\rho}_{CC}(t) | {\varPsi}_k^{\text{L}} \rangle \nonumber \nonumber \\
& = & \sum_\alpha \left\{{\varGamma}_{\alpha,jk} {\varLambda}_{\alpha,jk} + V_\alpha {\varGamma}_{\alpha,jk}\left[{\varPi}_{\alpha,jk}(t)+{\varPi}_{\alpha,kj}^*(t)\right]+V_\alpha^2{\varGamma}_{\alpha,jk}\ex^{-\im({\epsilon}_j-{\epsilon}_k^*)t}{\varOmega}_{\alpha,jk}\right\} , \label{eq:td1rdm}
\eeq
with
\beq
{\varGamma}_{\alpha,jk} & = & \langle {\varPsi}_j^{\text{L}} | \underline{\varGamma}_\alpha | {\varPsi}_k^{\text{L}} \rangle , \\ \nonumber \\
{\varLambda}_{\alpha,jk} & = & \frac{\im}{{\epsilon}_k^* - {\epsilon}_j}\left\{\frac{1}{\ex^{\beta({\epsilon}_k^*-\mu_\alpha)}+1} + \frac{1}{2\pi\im}\left[\psi\left(\frac{1}{2}-\frac{\beta({\epsilon}_k^*-\mu_\alpha)}{2\pi\im}\right)-\psi\left(\frac{1}{2}-\frac{\beta({\epsilon}_j-\mu_\alpha)}{2\pi\im}\right)\right]\right\} , \label{eq:lambda_finalresult} \nonumber \\ \\ 
{\varPi}_{\alpha,jk}(t) & = & \frac{\im}{({\eps}_k^* - {\eps}_j)({\eps}_k^* - {\eps}_j - V_\alpha)}\left\{\frac{\ex^{-\im({\eps}_j - {\eps}_k^*)t}}{\ex^{\beta({\eps}_k^* - \mu_\alpha)}+1} + \im\ex^{-\pi t /\beta}\ex^{-\im({\eps}_j - \mu_\alpha)t}\times\right.\nonumber \\
&  & \left. \left[\mathfrak{{F}}({\eps}_k^*-\mu_\alpha,t,\beta) + \frac{{\eps}_k^*-{\eps}_j - V_\alpha}{V_\alpha} \mathfrak{{F}}({\eps}_j-\mu_\alpha,t,\beta) - \frac{{\eps}_k^* - {\eps}_j}{V_\alpha} \mathfrak{{F}}({\eps}_j-\mu,t,\beta)\right]\right\} , \label{eq:pi_finalresult}\\ \nonumber \\ 
{\varOmega}_{\alpha,jk} & = & \frac{\frac{\im}{\ex^{\beta({\eps}_k^*-\mu)}+1}-\frac{1}{2\pi}\left[\psi\left(\frac{1}{2} - \frac{\beta({\eps}_j-\mu_\alpha)}{2\pi\im}\right) - \psi\left(\frac{1}{2} - \frac{\beta({\eps}_k^*-\mu)}{2\pi\im}\right)\right]}{({\eps}_k^*-{\eps}_j)({\eps}_k^* - {\eps}_j + V_\alpha)V_\alpha} \nonumber \\
& - & \frac{\frac{\im}{\ex^{\beta({\eps}_k^*-\mu_\alpha)}+1}-\frac{1}{2\pi}\left[\psi\left(\frac{1}{2} - \frac{\beta({\eps}_j-\mu)}{2\pi\im}\right) - \psi\left(\frac{1}{2} - \frac{\beta({\eps}_k^*-\mu_\alpha)}{2\pi\im}\right)\right]}{({\eps}_k^*-{\eps}_j)({\eps}_k^* - {\eps}_j - V_\alpha)V_\alpha} ,\label{eq:omega_finalresult}
\eeq
where $\beta$ is the inverse temperature, $\mu$ is the chemical potential and we also denoted the electro-chemical potential as $\mu_\a = \mu + V_\a$. In the above expressions $\psi$ is the digamma function\cite{digamma} and we defined another special function by $\mathfrak{{F}}(z,t,\beta) \equiv \frac{1}{\im\beta z + \pi} \ \hypf\left(1, \frac{1}{2}+\frac{\im\beta z}{2\pi}, \frac{3}{2}+\frac{\im\beta z}{2\pi}, \ex^{-2\pi t/\beta}\right)$ with $\hypf$ being the hypergeometric function\cite{hypf}.
Studying the asymptotic behaviour of the digamma and hypergeometric functions the results in Eqs.~\eqref{eq:lambda_finalresult}, \eqref{eq:pi_finalresult} and~\eqref{eq:omega_finalresult} can be shown to reduce to those in Ref.~\cite{Tuovinen2014} in the zero-temperature limit ($\beta\to\infty$)\cite{Tuovinen2015}.

\section{An introductory example}
\begin{figure}[tbp]
\centering
\begin{minipage}{0.375\textwidth}
\centering
\includegraphics[width=\textwidth]{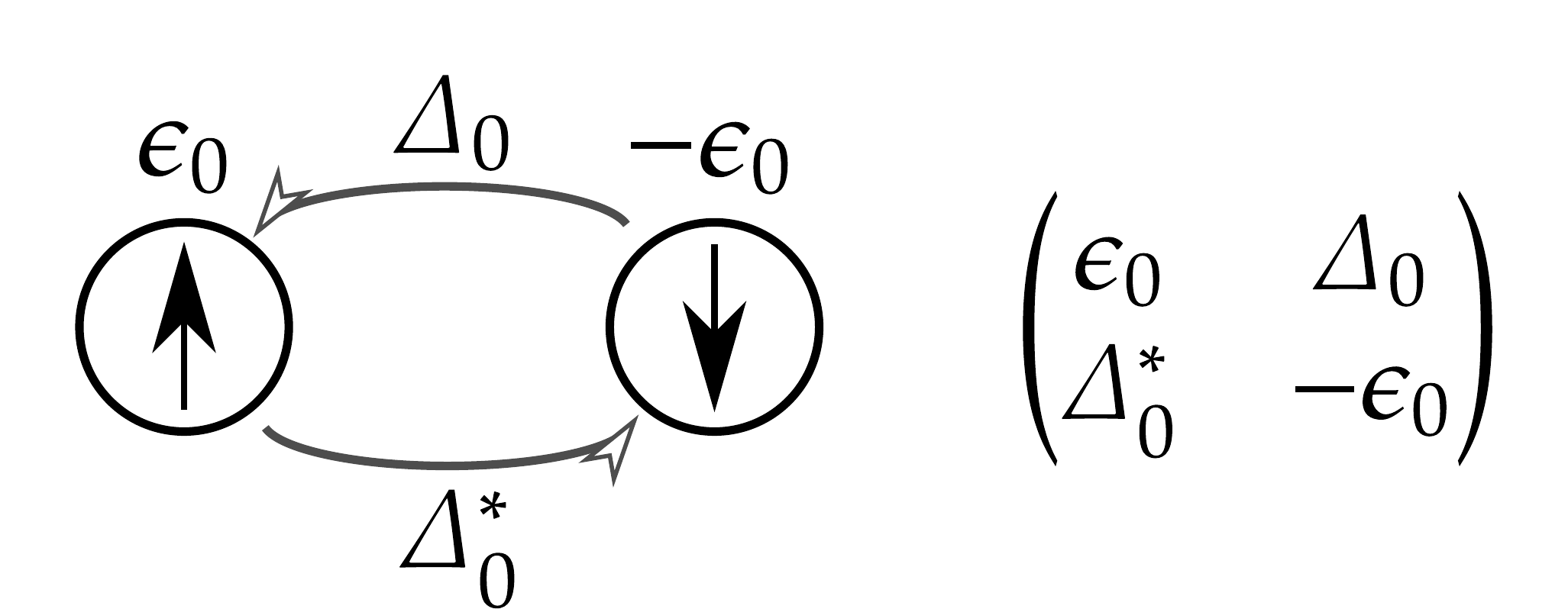}
\caption{\label{fig:dot}The dot viewed as a two-level system for different spins.}
\end{minipage}~~~~~
\begin{minipage}{0.375\textwidth}
\centering
\includegraphics[width=\textwidth]{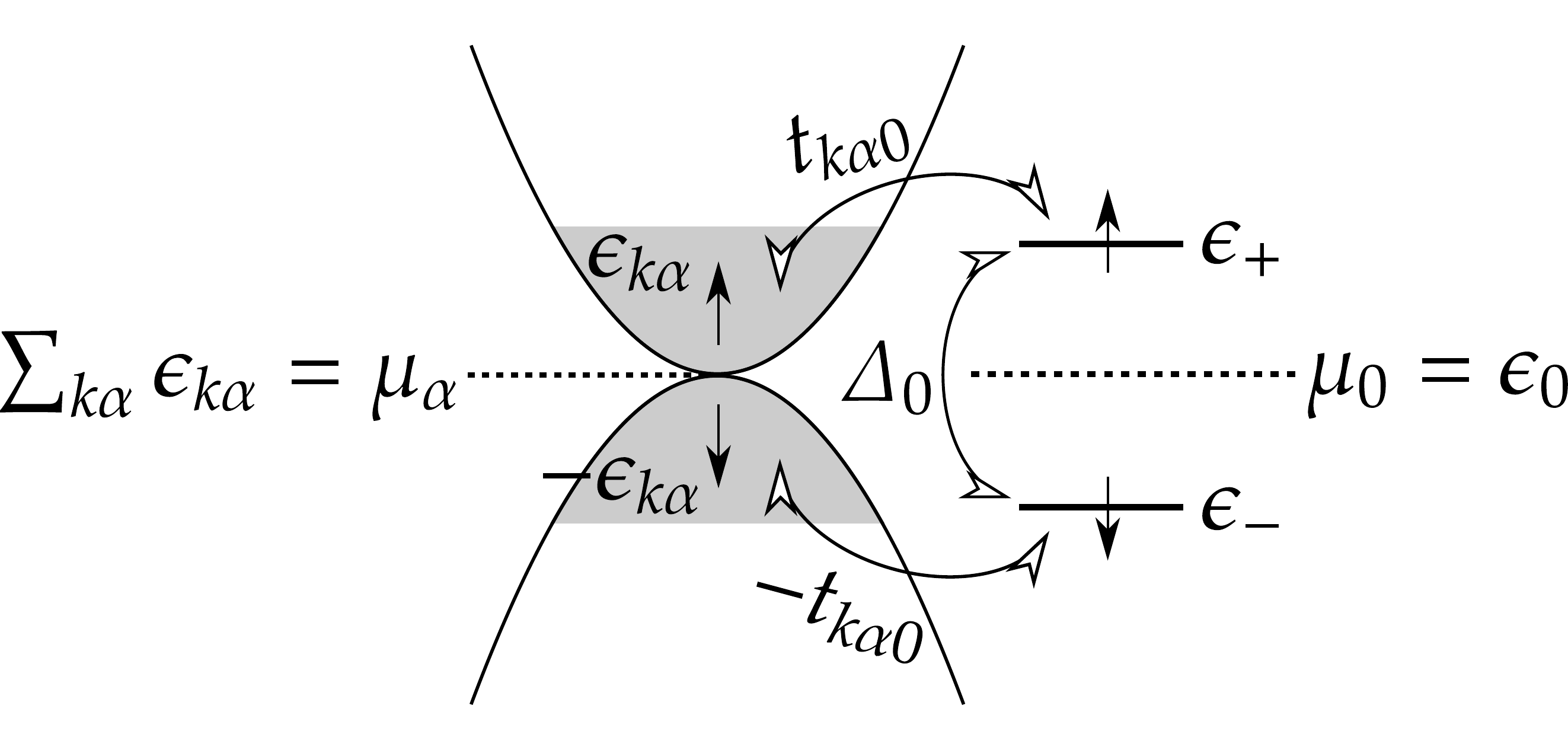}
\caption{\label{fig:nsn-diagram}Couplings between the lead and the dot in a transport setup.}
\end{minipage}
\end{figure}

Let us motivate the discussion for the NSN setup by means of a simple example. Consider a single dot connected to two leads for which the Hamiltonian can be separated in parts for leads, tunneling and dot, respectively as
\beq
\hat{H} & = & \sum_{k\a\sigma}\eps_{k\a}\hat{c}_{k\a\sigma}^\dagger \hat{c}_{k\a\sigma} + \sum_{k\a\sigma}t_{k\a 0}\hat{c}_{k\a\sigma}^\dagger \hat{c}_{0\sigma} + \sum_{k\a\sigma}t_{k\a 0}^* \hat{c}_{0\sigma}^\dagger\hat{c}_{k\a\sigma} + \eps_0\sum_{\sigma}\hat{c}_{0\sigma}^\dagger \hat{c}_{0\sigma} \nonumber \\
& + & \varDelta_0 \hat{c}_{0\uparrow}^\dagger \hat{c}_{0\downarrow}^\dagger + \varDelta_0^* \hat{c}_{0\downarrow} \hat{c}_{0\uparrow}
\eeq
with $\eps_{k\a}$ giving the level structure of the leads $\a\in\{L,R\}$, $t_{k\a 0}$ corresponding to the tunneling strength between the leads and the dot, and $\eps_0,\varDelta_0$ being the energy and the pairing strength in the dot, respectively. Let us introduce a new set of operators $\hat{\tilde{c}}_{x\sigma} = \hat{c}_{x\sigma}^\dagger$ obeying the fermionic anticommutation relation. The Hamiltonian can now be rewritten in terms of the new and old operators as
\beq\label{eq:nsn-example}
\hat{H} & = & \sum_{k\a}\eps_{k\a}\hat{c}_{k\a\uparrow}^\dagger \hat{c}_{k\a\uparrow} + \sum_{k\a}(-\eps_{k\a})\hat{\tilde{c}}_{k\a\downarrow}^\dagger\hat{\tilde{c}}_{k\a\downarrow} + \sum_{k\a}\eps_{k\a} \nonumber \\
& + & \sum_{k\a}\left(t_{k\a 0}\hat{c}_{k\a\uparrow}^\dagger \hat{c}_{0\uparrow} + t_{k\a 0}^*\hat{c}_{0\uparrow}^\dagger\hat{c}_{k\a\uparrow}\right) + \sum_{k\a}\left[(-t_{k\a 0})\hat{\tilde{c}}_{0\downarrow}^\dagger\hat{\tilde{c}}_{k\a\downarrow} + (-t_{k\a 0}^*)\hat{\tilde{c}}_{k\a\downarrow}^\dagger\hat{\tilde{c}}_{0\downarrow}\right] \nonumber \\
& + & \eps_0 \hat{c}_{0\uparrow}^\dagger\hat{c}_{0\uparrow} + (-\eps_0)\hat{\tilde{c}}_{0\downarrow}^\dagger\hat{\tilde{c}}_{0\downarrow} + \eps_0 + \varDelta_0\hat{c}_{0\uparrow}^\dagger \hat{\tilde{c}}_{0\downarrow} + \varDelta_0^* \hat{\tilde{c}}_{0\downarrow}^\dagger\hat{c}_{0\uparrow}\label{eq:old-new}
\eeq
where two constant shifts $\sum_{k\a}\eps_{k\a}$ and $\eps_0$ occur due to the anticommutation relations. Each term in Eq.~\eqref{eq:old-new} has a similar structure, $\hat{c}^\dagger\hat{c}$, and we may model the dot part as in Fig.~\ref{fig:dot}
where the matrix is of the form of Eq.~\eqref{eq:hc} and the corresponding eigenvalues are $\eps_{\pm} = \pm \sqrt{\eps_0^2 + |\varDelta_0|^2}$. The transport setup corresponding to Eq.~\eqref{eq:nsn-example} can then be viewed through the energy diagram in Fig.~\ref{fig:nsn-diagram} where we notice the nature of the constant shifts in Eq.~\eqref{eq:nsn-example}; they could be regarded as a chemical potential. The energy levels for the lead sector are raised so that the energy level continuum for the spin-up particles goes up from $\sum_{k\a}\eps_{k\a}$ and the energy level continuum for the spin-down particles goes down from $\sum_{k\a}\eps_{k\a}$. Similarly, for the dot sector we have the energy levels raised by $\eps_0$. The coupling terms $t_{k\a 0}$ connect separately the spin-up and spin-down particles between the leads and the dot, and the pairing strength term $\varDelta_0$ acts as a hopping term flipping the spins within the dot.

According to this picture for the NSN setup and the Nambu structure in Eq.~\eqref{eq:green2} we will evaluate the local bond currents for the more general structure in Eq.~\eqref{eq:hc} by
\be\label{eq:bondcurrent}
J_{mn}(t) = -\sum_\sigma\left[T_{mn}(G_{CC,\sigma}^<)_{nm}(t,t) + \hc\right]
\ee
and the Cooper pair density by
\be\label{eq:pairdensity}
P_m(t) = \im (F_{CC}^>)_{mm}(t,t)\ex^{2\im T_{mm}t}
\ee
satisfying the continuity equation\cite{Stefanucci2010}
\be\label{eq:contequ}
\frac{\ud}{\ud t} n_m(t) = \sum_n J_{mn}(t) - 4 \Im [\varDelta_m^* P_m(t)\ex^{-2\im T_{mm} t}] 
\ee
where the site density is the expectation value $n_m = \langle \hat{n}_m \rangle$ of $\hat{n}_m = \sum_\sigma \hat{c}_{m\sigma}^\dagger \hat{c}_{m\sigma}$. In the Nambu representation of the lesser Green's function the diagonal blocks therefore give rise to the bond current whereas the off-diagonal blocks correspond to the Cooper pair density. In the continuity equation~\eqref{eq:contequ} the two different terms on the right-hand side can also be identified as the normal current and the super current.

\section{TD response of a superconducting junction}\label{sec:results}
As a first study we show a numerical confirmation of the presented formula; in the limit when the superconducting gap $\varDelta$ and the temperature $1/\beta$ vanish we should recover equal results with the formula in Ref.~\cite{Tuovinen2014}. For the sake of simple interpretation of the transients let us take, as an example, a $2$-site tight-binding dimer coupled to two semi-infinite one-dimensional leads. Let the hopping parameter between the sites be equal everywhere: $t_\a = t_{\a C} = t_C \eqqcolon \epsilon_0$ (for $\a = L,R$) leading to the tunneling rate $\varGamma_\alpha = 2t_{\a C}^2/|t_\a| = 2\epsilon_0$. This parameter is the strength of the bandwidth matrix elements in the form of Eq.~\eqref{eq:sigmar}. In fact, WBA is not a very good approximation in this case, as the resonances are comparatively rather wide, but we are only comparing different formulae within the WBA, so the results should only be taken as comparative. Let us also bias the leads symmetrically to $V_L = -V_R = \epsilon_0$ with respect to the chemical potential $\mu=0$. We calculate, from the TD1RDM, the local bond current between the two sites in the dimer using Eq.~\eqref{eq:bondcurrent}. (Due to equal hopping parameters through the setup this is equal to the current through the lead interfaces modulo a minor time delay.) 

\begin{figure}
\centering
\includegraphics[width=0.85\textwidth]{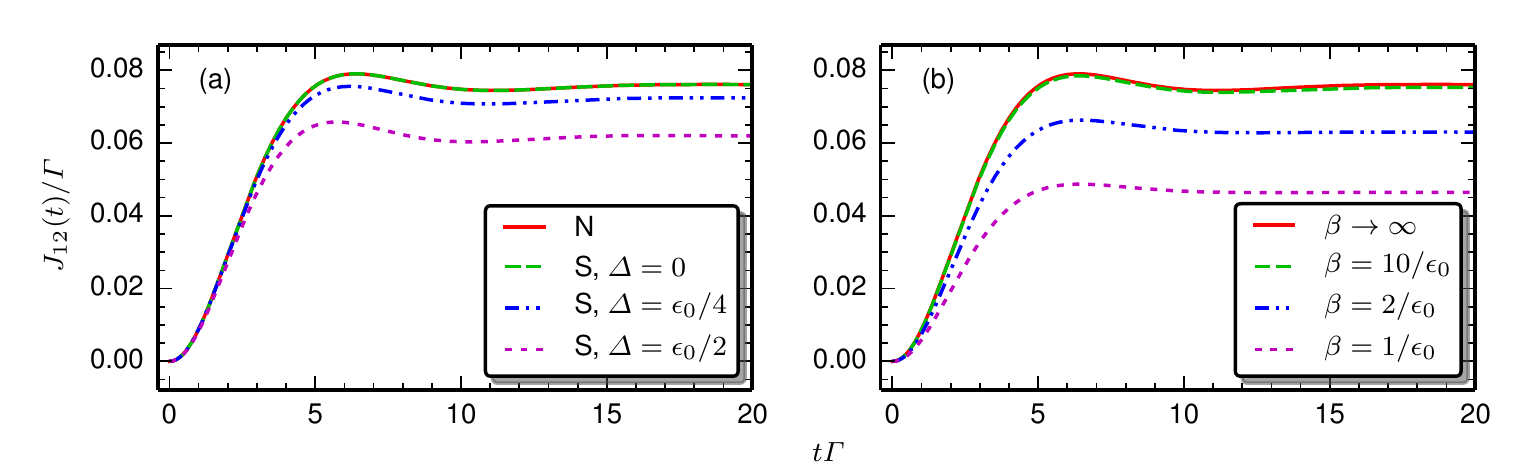}
\caption{(Color online) Transient current through a $2$-site dimer; comparison between different formulae and parameters. (a) Normal vs.\ superconducting central region at zero temperature; (b) normal central region at varying temperatures.}
\label{fig:comparison}
\end{figure}

In Fig.~\ref{fig:comparison}(a) we compare normal central region to a superconducting one by evaluating the TD1RDM from a normal Hamiltonian without the pairing field (non-Nambu), and from a Nambu Hamiltonian with varying pairing field strength at zero temperature. We see how the N and S ($\varDelta=0$) cases are on top of each other, and increasing the value for $\varDelta$ decreases the absolute value of the current through the central region as the energy levels of the central region get raised by $\sqrt{\varDelta^2 + \epsilon_0^2}$. In Fig.~\ref{fig:comparison}(b) we compare normal central regions at varying temperatures. In this comparative benchmark of varying temperature, we do not consider superconducting central regions as the temperature effects for the pairing field $\varDelta(T)$ should also be taken into account according to the self-consistent gap equation\cite{BCS2}. (In further simulations, also these parameters are considered in more detail.) The zero-temperature limit, $\beta \to \infty$, is evaluated from the results in Ref.~\cite{Tuovinen2014} which roughly agrees with an evaluation with $\beta=10/\epsilon_0$. (Increasing $\beta$ even more would naturally bring the curves exactly on top of each other.) Because the level structure of the studied system is symmetric around the chemical potential, increasing the temperature $1/\beta$ decreases the current due to broadening of the distribution function close to the Fermi level. In general, however, there is a possibility of enhancing the current by increasing the temperature if, for instance, the electronic levels were all above the Fermi level. In that case, the lead-states with energy higher than the energy of the levels get occupied, leading to an enhanced current.

\begin{figure}
\centering
\includegraphics[width=0.6\textwidth]{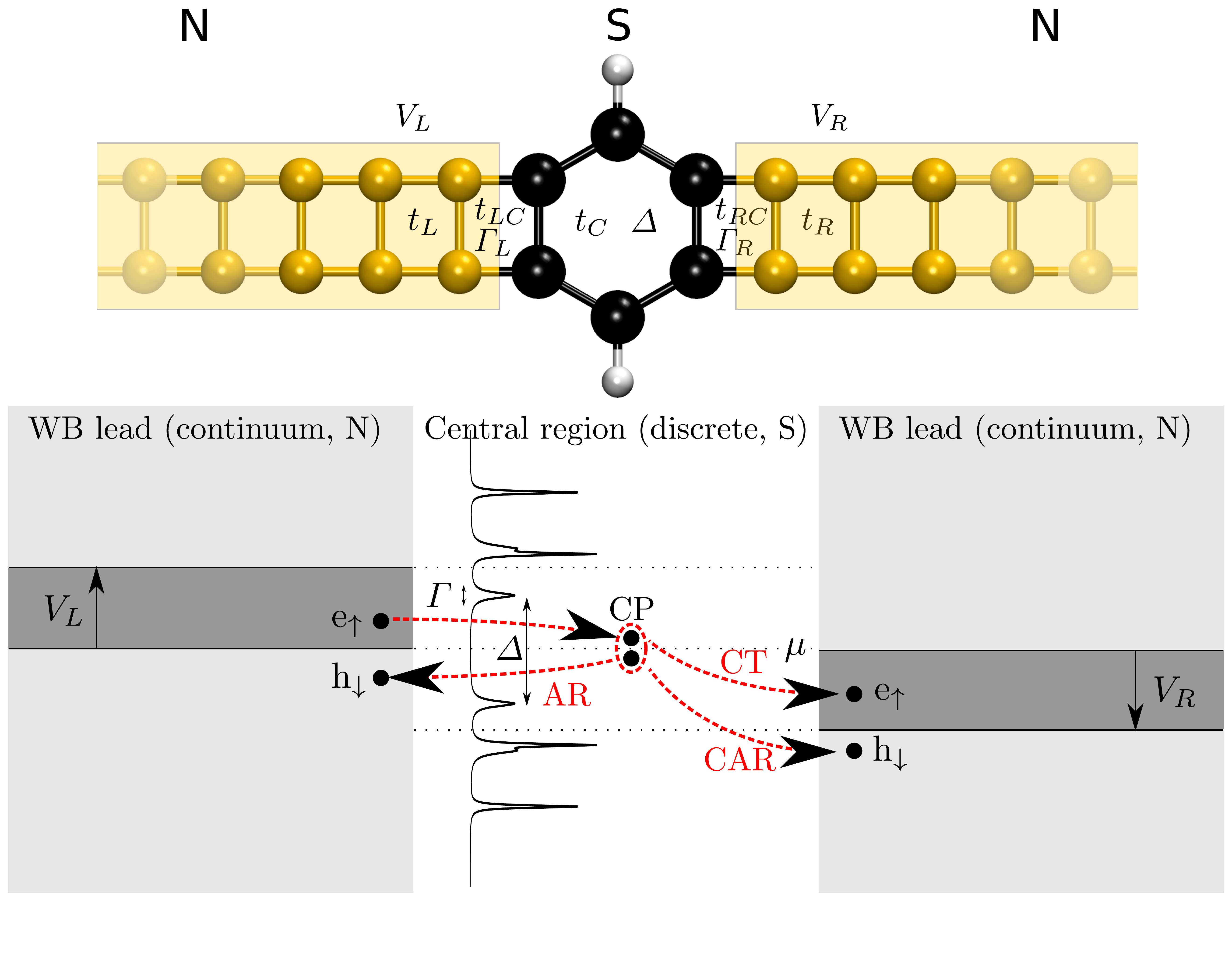}
\caption{(Color online) Transport setup in an NSN junction. Normal metal leads of continuum states undergo a level shift due to the bias voltage $V_{L/R}$ with respect to the chemical potential $\mu=E_{\text{F}}$ (Fermi level). The discrete level structure of the central region is determined by the tight-binding and gap $\varDelta$ parameters, and it is also broadened due to the coupling ($\varGamma$). Possible transition mechanisms are shown as CT, AR and CAR; see text for description.}
\label{fig:nsn-setup}
\end{figure}

After the comparisons presented above we can confidently conclude that the formulation of the NSN transport setup and the implementation of the formula for the TD1RDM is working properly. Now, we turn to a more concrete and physically relevant example, and we analyze the transient features in more detail. We consider a superconducting island made of a benzene-like molecule belonging to the class of quasi-one-dimensional polyacene chains\cite{Kivelson1983}. Transport in simple island setups has been studied, e.g., in Refs.~\cite{Averin1992,Maisi2013,Heimes2014,PhysRevLett.70.4138,PhysRevB.51.9084}, in a single-electron-tunneling level, where Coulomb blockade region is explored, and it is shown how the superconducting gap $\varDelta$ strongly and non-trivially affects the tunneling process. In polyacene samples (and in other carbon based materials, such as graphene) the superconductivity could be induced, e.g., by charge injection, chemical doping or using the proximity effect leading to critical temperatures ranging from $1$ to $10$~K\cite{Ojeda2009,Komatsu2012,Haberer2013,Yang2014}. Our setup is shown schematically in Fig.~\ref{fig:nsn-setup}. We model the benzene molecule in a single $\pi$-orbital tight-binding framework with the hopping parameter $t_C = -2{.}7$~eV\cite{Hancock2010}, and relate other energies to this scale. We also saturate the molecule's edges (longitudinally, in the transport direction) by hydrogen with modified tight-binding parameters for hydrogen on-site energies and hydrogen--carbon hopping\cite{Robinson2008}, respectively, so that there is no band gap in equilibrium. This condition is set because we want to isolate the effects from the superconducting gap $\varDelta$ without complicating the spectrum with the semiconducting gap. The coupling strength between the molecule and the leads and the lead hopping are chosen so that we are in weak coupling regime $\varGamma = 0{.}2$~eV.

Looking at the setup in Fig.~\ref{fig:nsn-setup} more closely suggests different transition mechanisms depending on the parameters. The simplest case is when the bias voltage $V_\alpha$ is larger than the superconducting gap $\varDelta$. In this case, all the levels inside the bias window act as transport channels, and transitions through the superconducting states are disrupted since the energy for the incoming electrons is high enough to break possible Cooper pairs (CP); this is referred to as normal tunneling (NT). If the bias voltage is smaller than the superconducting gap, this opens a possibility for the formation of a CP in the central region. In this case, it is possible to observe Andreev reflection (AR) between an electron and a hole in the source (or drain) lead forming the CP in the center, or to observe a crossed Andreev reflection (CAR) where an electron from the source (drain) lead is coupled to a hole in the drain (source) lead through the CP in the center. Also, direct tunneling of an electron via the CP, referred to as cotunneling (CT), is a possible transmission channel. Next, we simulate these different processes by a suitable parameter choice.

\begin{figure}[t!]
\centering
\begin{minipage}[t]{0.45\textwidth}
\centering
\includegraphics[width=\textwidth]{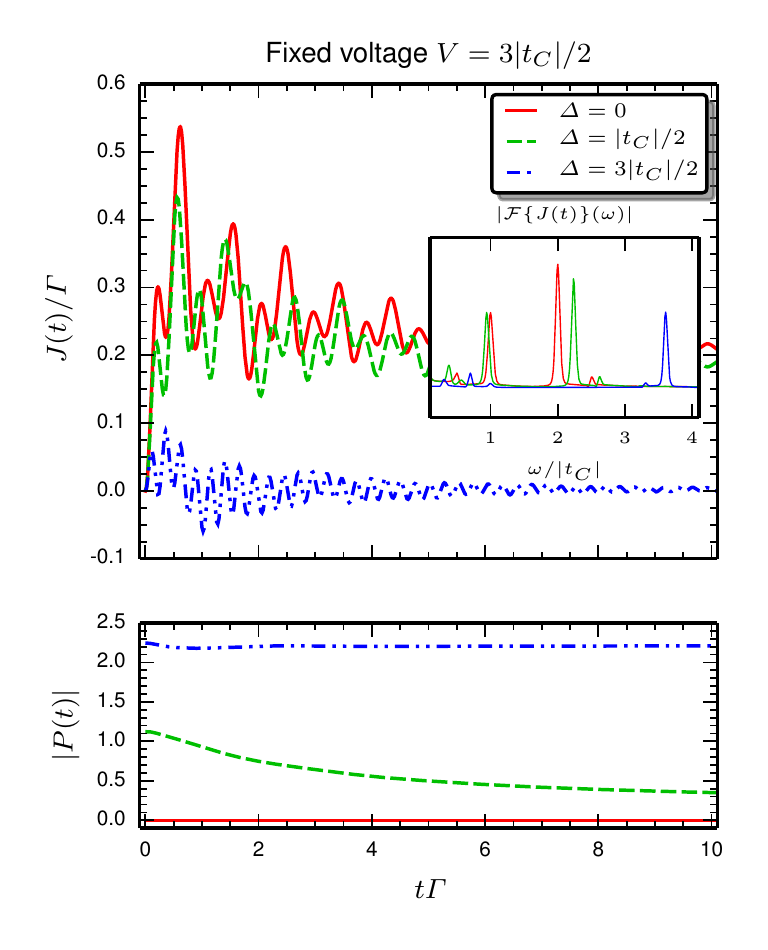}
\caption{(Color online) Transient currents (top panel) and pair densities (bottom panel) in the molecule when varying $\varDelta$. The inset shows the absolute value of the Fourier-transformed current.}
\label{fig:nt}
\end{minipage}\hspace{2pc}%
\begin{minipage}[t]{0.45\textwidth}
\centering
\includegraphics[width=\textwidth]{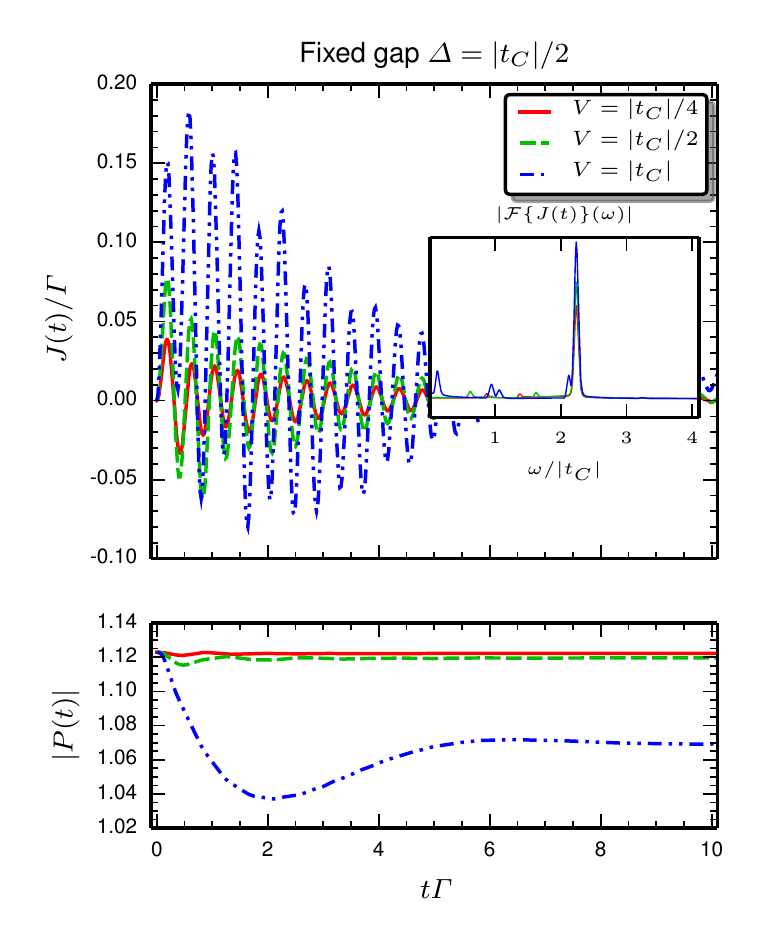}
\caption{(Color online) Transient currents (top panel) and pair densities (bottom panel) in the molecule when varying $V$. The inset shows the absolute value of the Fourier-transformed current.}
\label{fig:car}
\end{minipage}
\end{figure}

\begin{figure}[t!]
\centering
\begin{minipage}[t]{0.45\textwidth}
\centering
\includegraphics[width=\textwidth]{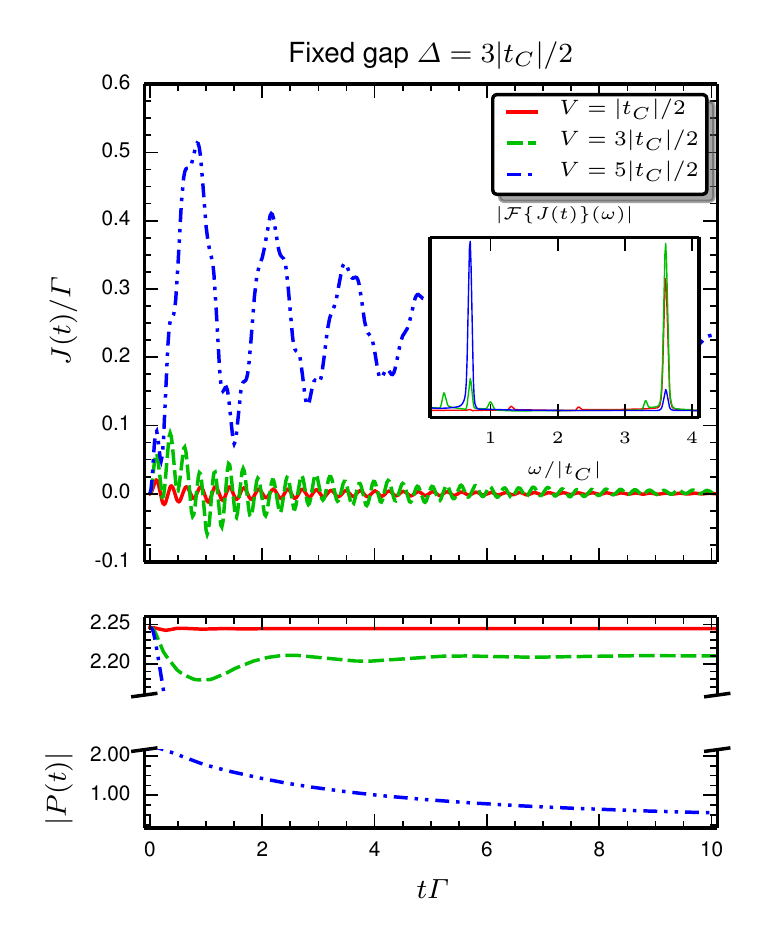}
\caption{(Color online) Transient currents (top panel) and pair densities (bottom panel) in the molecule when varying $V$. The inset shows the absolute value of the Fourier-transformed current.}
\label{fig:ar}
\end{minipage}\hspace{2pc}%
\begin{minipage}[t]{0.435\textwidth}
\centering
\includegraphics[width=\textwidth]{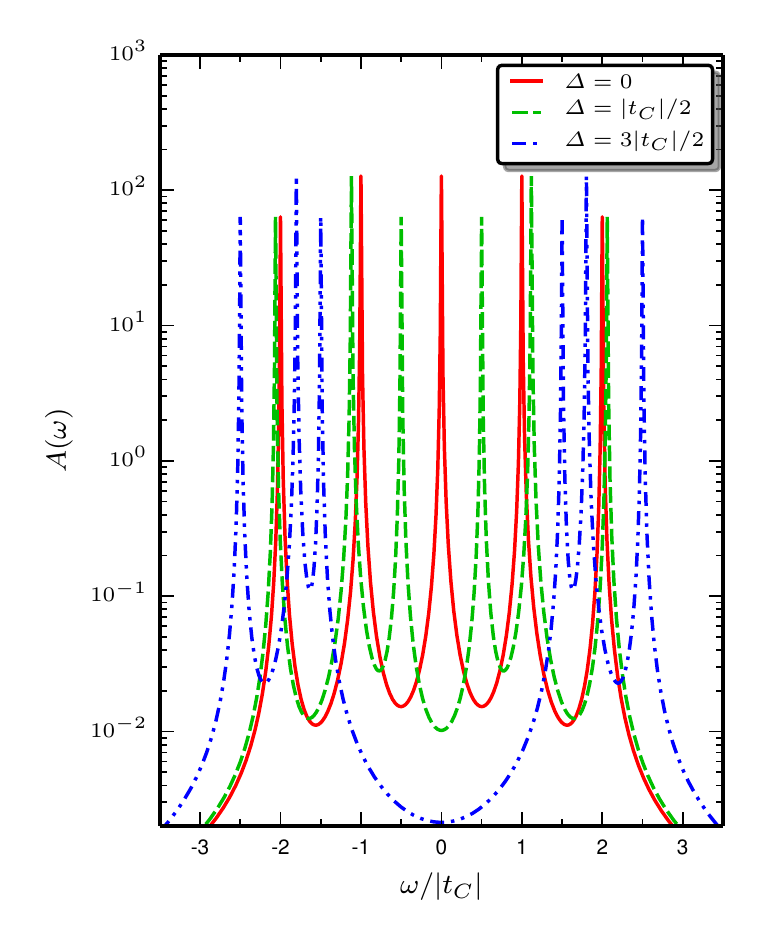}
\caption{(Color online) Spectral functions of the coupled benzene molecule when varying $\varDelta$.}
\label{fig:spec}
\end{minipage}
\end{figure}

We start with a familiar example by simulating NT: The condition is such that the bias window is larger than the gap. We will, in addition, fix the temperature, $\beta=100/|t_C|$, well below the critical temperature so that the gap can be approximated as the (constant) value at zero temperature $\varDelta(T=0)$. The sample is a benzene molecule consisting of $8$ atomic sites ($6$ carbon and $2$ hydrogen), coupled to the leads from four sites overall, see Fig.~\ref{fig:nsn-setup}. The bias voltage is symmetrically set to $V_L = -V_R = 3|t_C| / 2$ and the gap $\varDelta$ is varied but kept smaller than or equal to this value. The transient currents through the sample [calculated by summing the bond currents from Eq.~\eqref{eq:bondcurrent} transversally in the middle of the molecule], the corresponding Fourier transforms and the pair densities [calculated by summing the pair densities within the molecule: $P(t) = \sum_m P_m(t)$ from Eq.~\eqref{eq:pairdensity}] can be seen in Fig.~\ref{fig:nt}. Increasing the gap $\varDelta$ decreases the overall current as the conducting states are being pushed away from the bias window. This also leads to shifts in the transient frequencies seen in the Fourier spectrum. By looking also at the spectral function $A(\w) = -\frac{1}{\pi} \Im \Tr [G^{\text{R}}(\w)]$, where $G^{\text{R}}$ is the normal Nambu component of the retarded Green's function, plotted in Fig.~\ref{fig:spec} we can further identify the transitions.

In Fig.~\ref{fig:nt}, when $\varDelta=0$, we see two intramolecular transitions at frequencies $\w=|t_C|$ and $\w=2|t_C|$ which move a little when the gap is increased to $\varDelta=|t_C|/2$ corresponding to the shifted energy levels in the spectral function. We also observe two lead--molecule transitions at $\w=|t_C|/2$ and $\w=5|t_C|/2$ when $\varDelta=0$; these frequencies shift with the peaks in the spectral function corresponding to the fixed bias window at $V=3|t_C|/2$. The reason why we do not see a lead--molecule transition at $\w=3|t_C|/2$ when $\varDelta=0$, even though there is a zero-energy state in the molecule, is due to the fact that this state corresponds to the wavefunction's nodal planes being located exactly at the lead interface, and therefore it is an \emph{inert state} not taking part to the transient dynamics\cite{Tuovinen2013,Tuovinen2014}. Also, as the conditions are for NT, we observe the pair density within the molecule going to zero from its equilibrium value when $\varDelta < V$; this means that there are no out-of-equilibrium CPs forming in the central region, and we see no AR or CAR processes. When we set the gap equal to the bias window, we notice, first of all, that the steady-state current goes to zero since there are no transport channels within the bias window. Some transient oscillations are still present due to the states in the vicinity of the resonant window, which is seen as an intramolecular transition at $\w\sim 7|t_C|/2$.

Next, we will adjust the transport conditions to visualize the AR and CAR processes. We have the same benzene molecule as the central region at the same temperature, $\beta=100/|t_C|$. In Fig.~\ref{fig:car} we choose the gap as $\varDelta=|t_C|/2$, and in Fig.~\ref{fig:ar} we, on the other hand, choose the gap as $\varDelta=3|t_C|/2$. When $V\leq \varDelta$ we observe, in both cases, the transient current oscillating towards a zero steady-state current. The oscillation frequencies seen in the Fourier spectrum can be interpreted from the spectral function in Fig.~\ref{fig:spec} mainly as intramolecular transitions (around $\w=|t_C|$ and $\w=2|t_C|$). Interestingly, in Fig.~\ref{fig:car} also, when we increase the bias voltage above the superconducting gap $V=|t_C|$ there still are no other states within the bias window except the inert state split by $\varDelta$. This resonant window does not add anything to the transient dynamics (due to the inert state), but the static part of the density matrix is modified leading to nonzero steady-state current. This is interpreted as CT where, in addition to AR and CAR, also an electron is transferred from one lead to another via the CP. This is also confirmed by looking at the pair density as it remains nonzero also for $V=|t_C|$. With the larger gap in Fig.~\ref{fig:ar} and for $V\leq\varDelta$ we observe CP formation within the molecule but for $V>\varDelta$ the pair density goes to zero. For the smaller voltages we mainly find the first intramolecular transition at around $\w=7|t_C|/2$. For larger voltages we also see the lead--molecule transitions at lower frequencies, and we recover again the NT regime as the bias voltage is high enough for breaking the CPs within the molecule.

In this transport setup, it is not, in general, easy to distinguish between AR and CAR processes as they involve multiple steps. One possible case is when an electron hops from the lead to the central region (molecule--lead transition), then ``transfers'' from the spin-up sector in the central region to the spin-down sector for which the probability is given by the pairing strength (``intramolecular'' transition between the two branches of states split by $\varDelta$), and then finally hops back to the lead as a hole (molecule--lead transition). As all these transitions are visible in the transient oscillations, we may only conclude whether AR and CAR processes are present or not.

\section{Conclusions and outlook}\label{sec:concl}

We presented an extension to the time-dependent Landauer--Büttiker formalism, discussed in Refs.~\cite{svlbook,Tuovinen2013,Tuovinen2014}, to include superconducting central region in the transport setup, and to evaluate the TD1RDM at arbitrary temperatures. The derived formulae are analytic and closed expressions involving known special functions, and they can readily be implemented to study various quantum transport problems very efficiently and also at large temporal and spatial scales.

As an application of the presented formalism we simulated transport in a superconducting benzene-like molecule attached to two-dimensional normal metal leads. Assigning a proper parameter set for the transport window and the superconducting gap, we observed formation of Cooper pairs within the central molecule leading to Andreev reflection processes.

\ack
R.T. thanks the Väisälä Foundation of The Finnish Academy of Science and Letters for financial support. R.v.L. thanks the Academy of Finland for support. E.P. and G.S. acknowledge funding by MIUR FIRB Grant No. RBFR12SW0J.

\bibliography{refs}

\end{document}